\documentclass{article}

\usepackage{rotating}
\usepackage{url}
\usepackage{subcaption}
\usepackage{hyperref}
\usepackage{multirow}

\title{\bfseries Revisiting the Futamura Projections:\\ A Diagrammatic Approach}

\author{{\bfseries Brandon M. Williams} and {\bfseries Saverio Perugini}\\
\normalsize Department of Computer Science\\
\normalsize University of Dayton\\
\normalsize 300 College Park\\
\normalsize Dayton, Ohio\ \ 45469--2160\ \ USA\\
\normalsize Tel: +001 (937) 229--4079, Fax: +001 (937) 229--2193\\
\normalsize E-mail: \url{BrandonWilliamsCS@gmail.com},~\url{saverio@udayton.edu}\\
\normalsize WWW: \url{http://academic.udayton.edu/SaverioPerugini}}

\newcommand{\mix}{mix}
\newcommand{\funceval}[1]{[\![#1]\!]}
\newcommand{\pattident}[1]{#1} 
\newcommand{\pattlang}[1]{\mathcal{#1}} 
\newcommand{\instident}[1]{\mathtt{#1}} 
\newcommand{\instlang}[1]{\mathtt{#1}} 
\newcommand{\machlang}[1]{\texttt{#1}} 

\begin{document}
\sloppy

\maketitle
\thispagestyle{empty}


\begin{abstract}
    The advent of language implementation tools such as
	\textit{PyPy} and \textit{Truffle}/\textit{Graal}
	have reinvigorated and broadened interest in topics related to automatic compiler
	generation and optimization.  Given this broader interest, we revisit the
	\textit{Futamura Projections} using a novel diagram scheme.   Through these diagrams
	we emphasize the recurring patterns in the Futamura Projections while addressing
	their complexity and abstract nature.  We anticipate that this approach will
	improve the accessibility of the Futamura Projections and help foster analysis
	of those new tools through the lens of partial evaluation.
\end{abstract}

	\textit{Graal};
	interpretation;
	partial evaluation;
	program transformation;
	\textit{PyPy};
	\textit{Truffle}

\section{Introduction}

The \textit{Futamura Projections} are a series of program signatures reported
by~\cite{partialEvaluationComputationProcess} (a reprinting
of~\cite{partialEvaluationComputationProcessOrig}) designed to create a program
that generates compilers. This is accomplished by repeated applications of a
\textit{partial evaluator} that iteratively abstract away aspects of the
program execution process. A partial evaluator transforms a program given any
subset of its input to produce a version of the program that has been
specialized to that input.  The partial evaluation operation is referred to as
\textit{mixed computation} because partial evaluation involves a mixture of
interpretation and code generation~\cite{ershovPE}.  In this paper, we will
provide an overview of typical program processing and discuss the Futamura
Projections. We apply to the Projections a novel diagram scheme which
emphasizes the relationships between programs involved in the projections. We
also discuss related topics in the context of the Futamura Projections through
application of the same diagram scheme.

\begin{figure}
\centering
\includegraphics[scale=0.75]{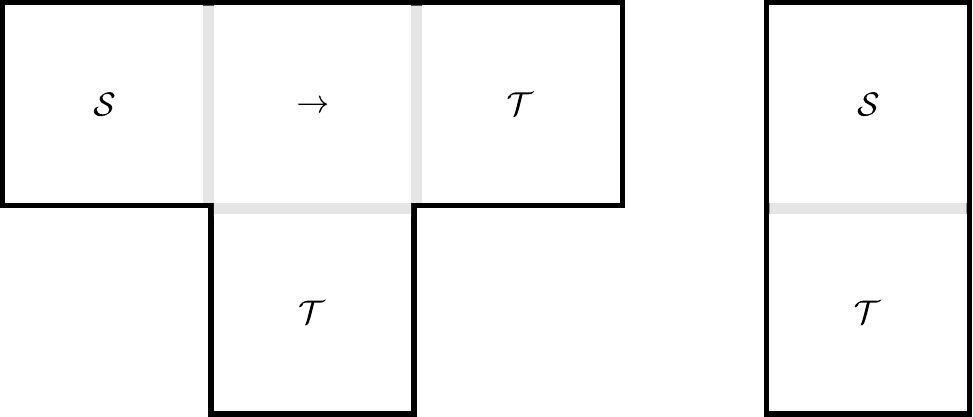}
\caption{The classical $\mathcal{T}$ and $\mathcal{I}$ diagrams, referring to the shapes
of the diagrams~\cite{JonesBook}.}
\label{fig:TIDiagramsExample}
\end{figure}

A common syntax for the graphical presentation of program interpretation and
transformation involves the classical $\mathcal{I}$ (for interpreter) and
$\mathcal{T}$ (for translator, meaning compiler) diagrams, referring to the
shapes of the diagrams, respectively~\cite{JonesBook}.
Fig.~\ref{fig:TIDiagramsExample} illustrates the syntax of each diagram.  An
$\mathcal{I}$ diagram specifies the interpreted language spatially above the
implementation language. A $\mathcal{T}$ diagram represents a compiler with
an arrow from its source to target languages spatially above its implementation
language.  These diagrams are simple to draw and recognize, and excel at
showing how multiple languages relate during compilation and interpretation.
However, although the diagrams can be extended to express a partial evaluator
(as described in~\cite{fourthProj}), they are not especially suited to
representing the partial evaluation process due to the secondary input given to
the partial evaluator. Additionally, the many languages and programs involved
in the Futamura Projections add complexity that is not addressed by this style
of diagram.  In contrast, we believe our diagrams make clear the relationships
between the various languages and programs involved in the Futamura Projections
and hope that they will improve the accessibility of the Projections.

\section{Programs Processing Other Programs}
\label{sec:Processing}

\begin{table}
\centering
\caption{Legend of symbols and terms used in 
$\S$~\ref{sec:Processing} and $\S$~\ref{sec:Futamura}.}
    \resizebox{\textwidth}{!}{
    \begin{tabular}{|l|l|l|}
    \hline
    \multicolumn{1}{|c}{\textbf{Symbol}} & \multicolumn{1}{|c|}{\textbf{Example}} & \multicolumn{1}{|c|}{\textbf{Description}} \\
    \hline
    $\pattident{program}$ & $\instident{pow}$ & A miscellaneous program. \\
    $\pattident{p}_{\pattident{n}}$ & $\instident{b}$ & A parameter. \\
    $\pattident{a}_{\pattident{n}}$ & $\instident{3}$ & An argument. \\
    $\pattident{compiler}^{\pattlang{S}\rightarrow{}\pattlang{T}}_{\pattlang{T}}$ & $\instident{compiler^{\instlang{C}\rightarrow{}\instlang{x86}}_{x86}}$ & A compiler from language $\pattlang{S}$ to language $\pattlang{T}$, implemented in $\pattlang{T}$. \\
    $\pattident{program}_{\pattlang{L}}$ & $\instident{pow.c}$ & A miscellaneous program implemented in language $\pattlang{L}$. \\
    $interpreter^{\pattlang{S}}_{\pattlang{T}}$ & $\instident{interpreter}^{\instlang{C}}_{\instlang{x86}}$ & An interpreter for language $\pattlang{S}$ implemented in language $\pattlang{T}$. \\
    $\pattident{partial\ input_{static}}$ & $\instident{partial\ input_{static}}$ & A subset of input for a program being specialized by $\pattident{\mix{}}$. \\
    $\pattident{program^{\prime}}_\pattlang{T}$ & $\instident{square}_{\instlang{x86}}$ & A specialized program implemented in language $\pattlang{T}$. \\
    $\pattident{\mix{}}_{\pattlang{T}}$ & $\instident{\mix{}_{\instlang{x86}}}$ &A partial evaluator implemented in language $\pattlang{T}$. \\
    [0.75ex]
    $\pattident{compiler}$ & $\instident{compiler}$ & \multirow{2}{*}{A compiler generator implemented in language $\pattlang{T}$.} \\
    $\pattident{generator}_{\pattlang{T}}$ & $\instident{generator}_{\instident{x86}}$ & \\
    \hline
    \end{tabular}}
    \label{tab:legend}
\end{table}

\begin{figure}[t]
    \centering
    \begin{subfigure}{\textwidth}
        \centering
        \includegraphics[scale=0.85]{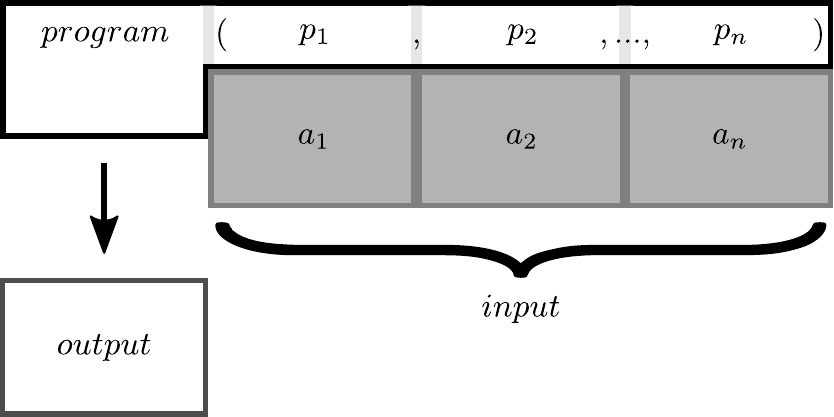}
        \caption{Program execution depicted as a machine.}
        \label{fig:BasicPattern}
    \end{subfigure}
    \begin{subfigure}{\textwidth}
        \centering
        \includegraphics[scale=0.85]{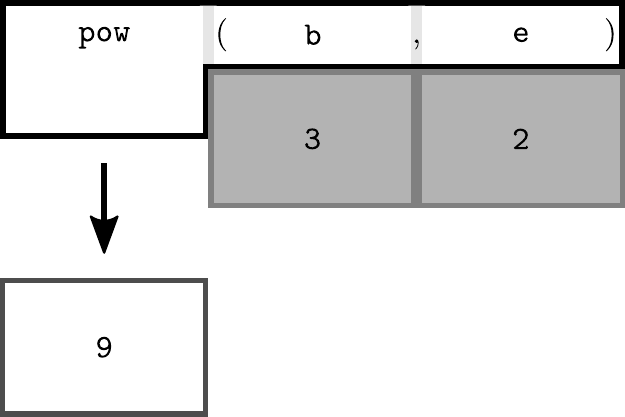}
        \caption{Program execution instance.}
        \label{fig:BasicExample}
    \end{subfigure}
\caption{Program execution.}
\end{figure}

We use notation in this paper for partial evaluation and associated programming
language concepts from~\cite{JonesBook,introPartialEvaluation}.  That notation
is commonly used in papers on partial
evaluation~\cite{fourthProj,cogenSixLines}.  In particular, we use
$\funceval{-}$ to denote a \textit{semantics function} which represents the
evaluation of a program by mapping that program's input to its output.  For
instance, $\funceval{p}[s,d]$ represents the program $p$ applied to the inputs
$s$ and $d$.  We use the symbol \texttt{mix} to denote the partial evaluation
operation~\cite{JonesBook,introPartialEvaluation}, which involves a
\texttt{mix}ture of interpretation and code generation.  Table~\ref{tab:legend}
is a legend mapping additional terms and symbols used in this article to their
description.

Program execution can be represented equationally as
$\funceval{\pattident{program}}[\pattident{a_{1}}, \pattident{a_{2}}, ...,
\pattident{a_{n}}] = [\pattident{output}]$~\cite{JonesBook}.  Alternatively,
the diagram in Fig.~\ref{fig:BasicPattern} depicts a program as a machine that
takes a collection of input boxes, marked by divided \textit{slots} of the
input \textit{bar}, and produces an output box.  We use this diagram syntax to
aid in the presentation of complex relationships between programs, inputs, and
programs treated as inputs (i.e., data).  Each input area corresponds to part
of a C-function-style signature that names and positions the inputs. The input
is presented in gray to distinguish it from the program and its input bar.
Fig.~\ref{fig:BasicExample} shows this pattern applied to a program that takes
a base $\instident{b}$ and an exponent $\instident{e}$ and raises the base to
the power of the exponent. In this case, $\instident{3}$ raised to the power of
$\instident{2}$ produces $\instident{9}$, or
$\funceval{\instident{pow}}[\instident{3}, \instident{2}] = [\pattident{9}]$.

\subsection{Compilation}

\begin{figure}
    \centering
    \begin{subfigure}{\textwidth}
        \centering
        \includegraphics[scale=0.75]{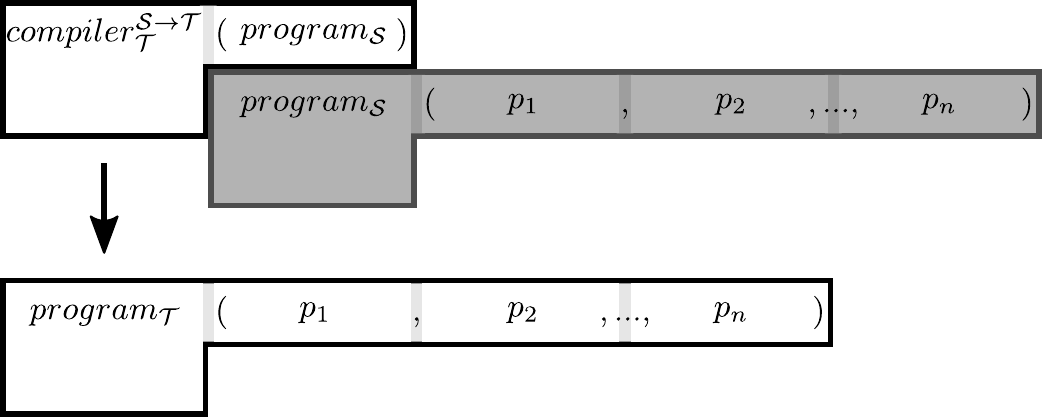}
        \caption{The compilation process.}
        \label{fig:CompilerPattern}
    \end{subfigure}
    \begin{subfigure}{\textwidth}
        \centering
        \includegraphics[scale=0.75]{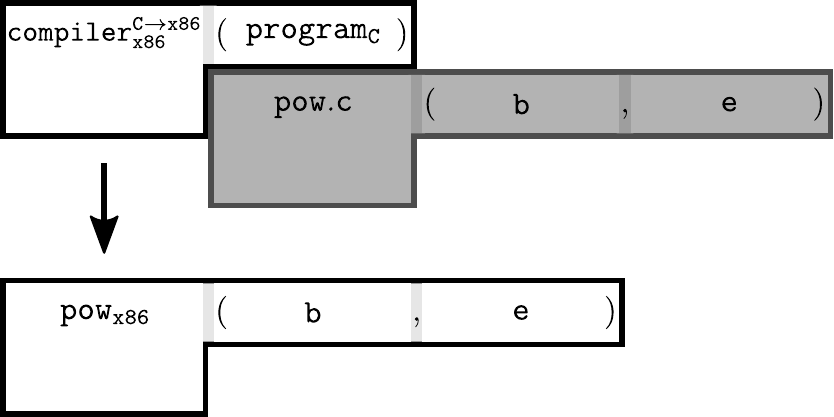}
        \caption{Compilation example.}
        \label{fig:CompilerExample}
    \end{subfigure}
    \begin{subfigure}{\textwidth}
        \centering
        \includegraphics[scale=0.75]{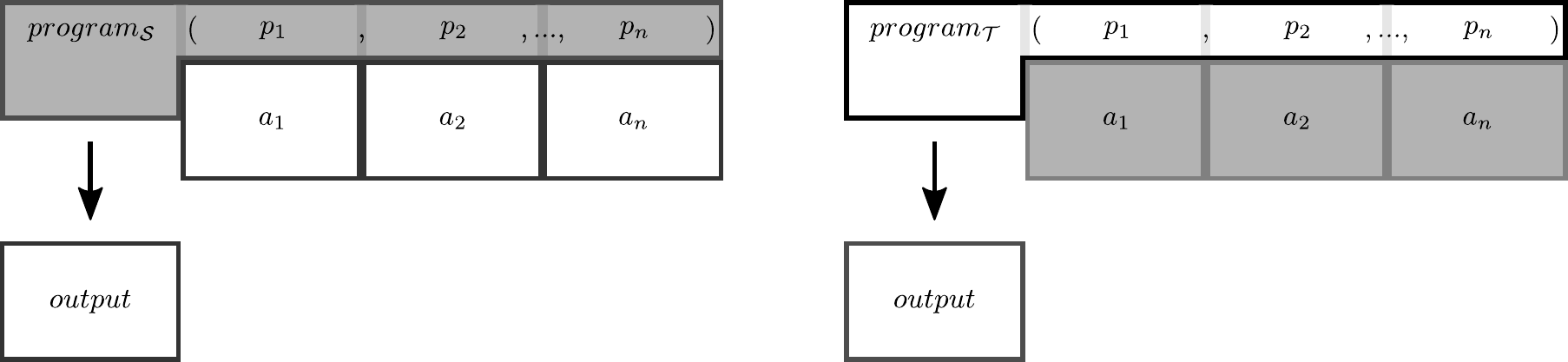}
        \caption{Comparison of compiler input and output.}
        \label{fig:CompilerPatternOutput}
    \end{subfigure}
    \begin{subfigure}{\textwidth}
        \centering
        \includegraphics[scale=0.75]{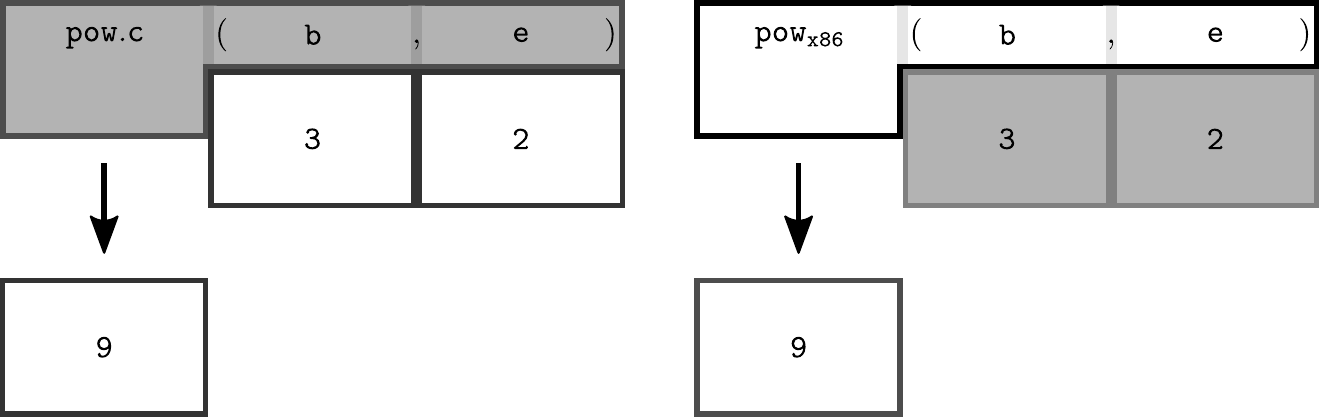}
        \caption{Comparison of compiler example input and output.}
        \label{fig:CompilerExampleOutput}
    \end{subfigure}
    \caption{Execution through compilation.}
\end{figure}

Programs written in higher-level programming languages such as C must be either compiled to a natively runnable language (e.g., the \machlang{x86} machine language) or evaluated by an interpreter. A compiler is simply a program that translates a program from its source language to a target language.
This process is described equationally as
$\funceval{\pattident{compiler}^{\pattlang{S}\rightarrow{}\pattlang{T}}_{\pattlang{T}}}[\pattident{program}_{\pattlang{S}}] = [\pattident{program}_{\pattlang{T}}]$,
or diagrammatically in Fig.~\ref{fig:CompilerPattern}. For clarity, the implementation language appears as a subscript of any program name.
Compilers will also have a superscript with an arrow from the source (input) language to the target (output) language.
If language $\pattlang{T}$ is natively executable, both the depicted compiler and its output program are natively executable. If $\instident{pow}$ from Fig.~\ref{fig:BasicExample} is written in C, it can be compiled to the \machlang{x86} machine language with a compiler as depicted in Fig.~\ref{fig:CompilerExample} or expressed equationally as
$\funceval{\instident{compiler}^{\instlang{C}\rightarrow{}\instlang{x86}}_{\instlang{x86}}}[\instident{pow.c}] = [\instident{pow}_{\instlang{x86}}]$.
Figs.~\ref{fig:CompilerPatternOutput} and \ref{fig:CompilerExampleOutput} illustrate that the source and the target programs are semantically equivalent.

\subsection{Interpretation}

\begin{figure}
\centering
\begin{subfigure}{\textwidth}
	\centering
	\includegraphics[scale=0.75]{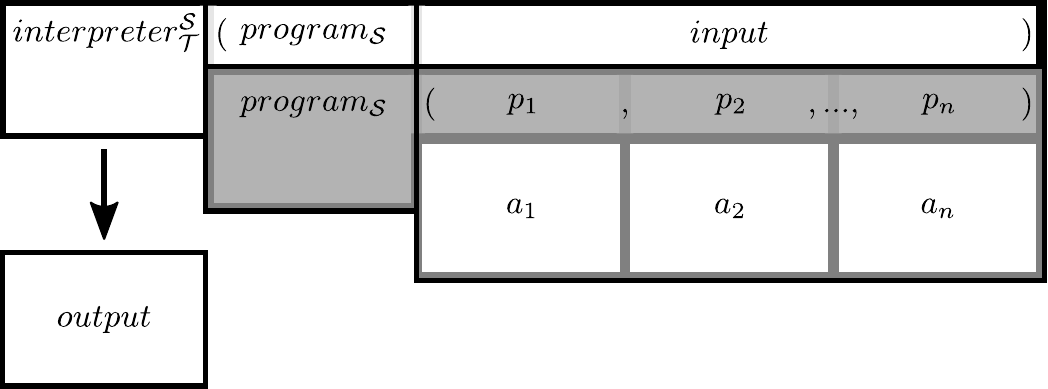}
	\caption{The interpretation process.}
	\label{fig:InterpreterPattern}
\end{subfigure}
\begin{subfigure}{\textwidth}
	\centering
	\includegraphics[scale=0.75]{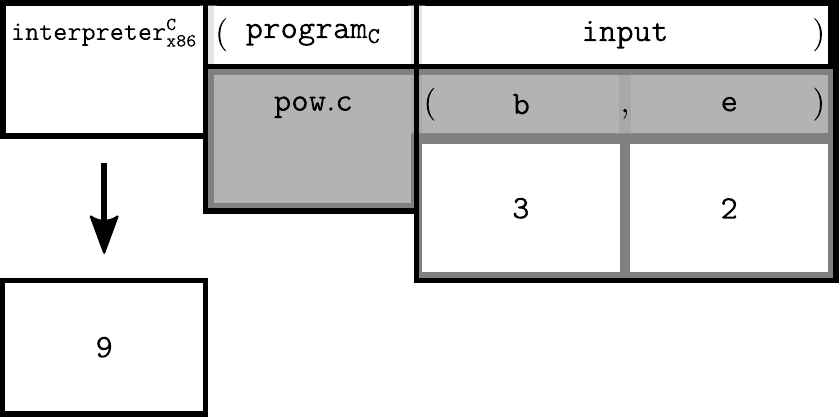}
	\caption{Instance of program interpretation.}
	\label{fig:InterpreterExample}
\end{subfigure}
\caption{Execution by interpretation.}
\end{figure}

The gap between a high-level source language and a natively executable target
language can also be bridged with the use of an interpreter. ``The interpreter
for a computer language is just another program'' implemented in the target
language that evaluates the program given the program's input and producing its
output~\cite{EOPL3}. The interpretation pattern, described equationally as
$\funceval{\pattident{interpreter}^{\pattlang{S}}_{\pattlang{T}}}[\pattident{program}_{\pattlang{S}},
\pattident{input}] = [\pattident{output}]$, is depicted in
Fig.~\ref{fig:InterpreterPattern}. The interpreter has the previously
established implementation language subscript, with a superscript indicating
the interpreted language. The input program's input bar extends into the
interpreter's next input slot, which serves to indicate which individual input
is associated with each of the program's own input slots.  However, as the
inputs are actually being provided directly to the interpreter, an outline is
drawn around each input to the program being executed. By convention, the
background shading is alternated to differentiate inputs, while the borders of
inputs remain gray. This pattern is applied to the $\instident{pow.c}$ program
in Fig.~\ref{fig:InterpreterExample}; the C program is being executed by an
interpreter implemented in \machlang{x86} to produce the output from
$\instident{pow.c}$ given its input.  This interpretation is represented
equationally as
$\funceval{\instident{interpreter}^{\instlang{C}}_{\instlang{x86}}}[\instident{pow.c},
\instident{3}, \instident{2}] = [\instident{9}]$.

\subsection{Partial Evaluation}

\begin{figure}
    \centering
    \begin{subfigure}{\textwidth}
        \centering
        \includegraphics[scale=0.75]{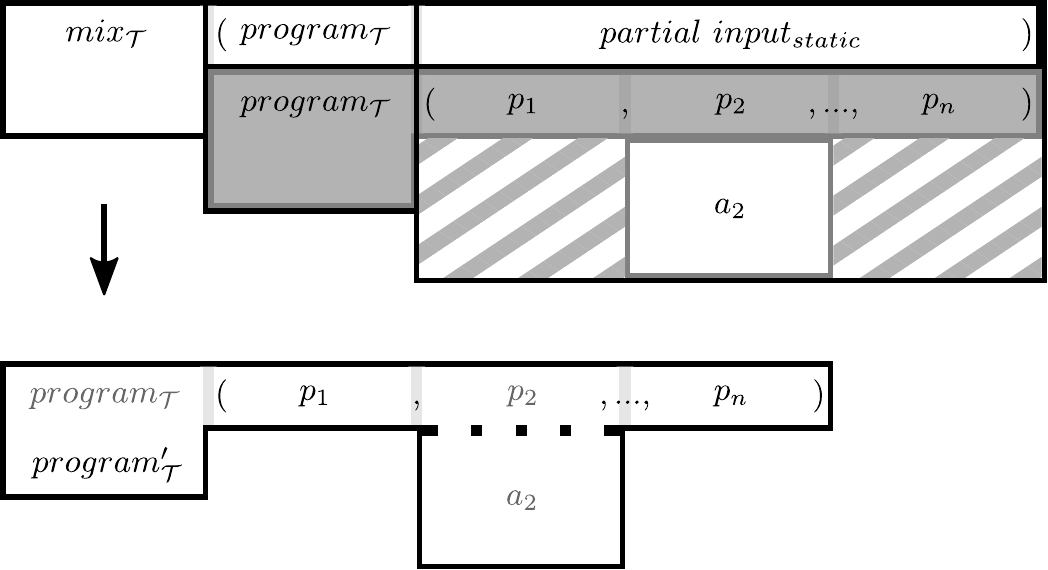}
        \caption{The partial evaluation process.}
        \label{fig:MixPattern}
    \end{subfigure}
    \begin{subfigure}{\textwidth}
        \centering
        \includegraphics[scale=0.75]{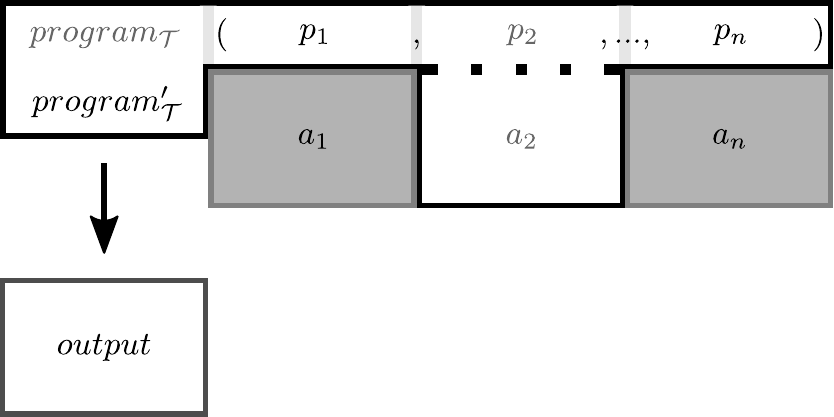}
        \caption{Partial evaluation output.}
        \label{fig:MixPatternOutput}
    \end{subfigure}
    \begin{subfigure}{\textwidth}
        \centering
        \includegraphics[scale=0.75]{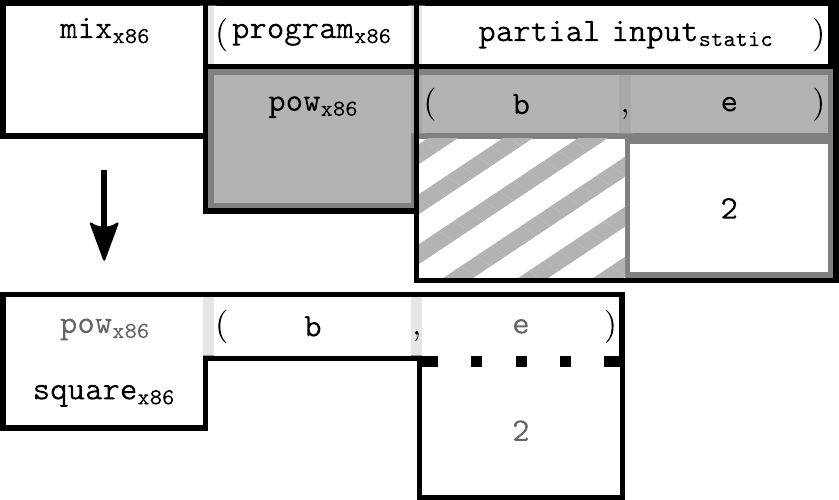}
        \caption{Instance of partial evaluation.}
        \label{fig:MixExample}
    \end{subfigure}
    \begin{subfigure}{\textwidth}
        \centering
        \includegraphics[scale=0.75]{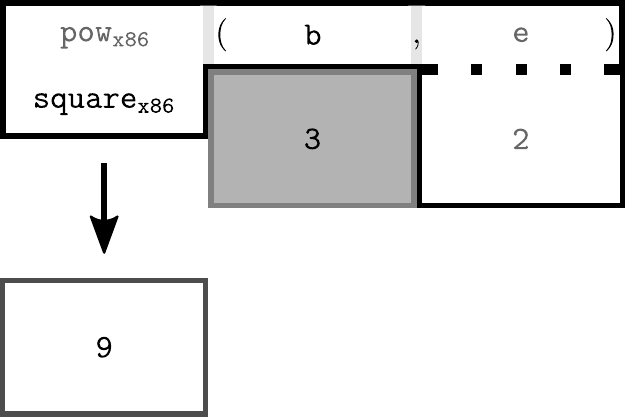}
        \caption{Output of partial evaluation instance.}
        \label{fig:MixExampleOutput}
    \end{subfigure}
    \caption{Partial evaluation.}
\end{figure}

With typical program evaluation, complete input is provided during program
execution.  With partial evaluation, on the other hand, partial
input---referred to as \textit{static} input---is given in advance of program
execution. This partial input is given to $\instident{\mix{}}$ with the program
to be evaluated and \textit{specializes} the program to the input.
The specialized program then accepts only
the remaining input---referred to as the \textit{dynamic} input---and produces
the same output as would have been produced by evaluating the original program
with complete input.  For reasons explained below, the Futamura Projections
require that $\instident{\mix{}}$ be implemented in the same language as the
program it takes as input; diagrams including $\instident{\mix{}}$ provide a
subscript that represents the implementation language of $\instident{\mix{}}$
as well as the that of the input and output programs.

Partial evaluation is described equationally as
$\funceval{\pattident{\mix{}}_{\pattlang{T}}}[\pattident{program}_{\pattlang{T}},
\pattident{partial\ input_{static}}] =
[\pattident{program^{\prime}}_{\pattlang{T}}]$ and depicted in
Fig.~\ref{fig:MixPattern}.  Here, a program is being passed to
$\instident{\mix{}}$ with partial input---specifically, only its second
argument. The result is a transformed version of the program specialized to the
input; the input has been propagated into the original program to produce a new
program. This specialized transformation of the input program is called a
\textit{residual program}~\cite{ershovPE}.  Notice how the shape of the
residual program matches the shape of the input program combined with the
static input.  Notice also in Fig.~\ref{fig:MixPatternOutput} that the shape of
the program combined with the remainder of its input matches the shape of the
typical program execution shown in Fig.~\ref{fig:BasicPattern}.  However, the
input has been visually fused to the program, represented by the dotted line.
In addition, the labels for the original program, the second input slot, and
the static input have been shaded gray; while the residual program is entirely
comprised of these two components, its input interface has been modified to
exclude them.  In other words, while the semantics of the components are still
present, they are no longer separate entities.  The equational representation
of this residual program shows the simplicity of its behavior:
$\funceval{\pattident{program^{\prime}}_{\pattlang{T}}}[\pattident{a_{1}},
\pattident{a_{3}}, ..., \pattident{a_{n}}] = [\pattident{output}]$.

The partial evaluation pattern is applied to the $\instident{pow_{x86}}$
program in Fig.~\ref{fig:MixExample}.  If $\instident{pow_{x86}}$ is partially
evaluated with static input $\instident{e}\!=\!\instident{2}$, the result is a
power program that can only raise a base to the power 2. This example is
represented equationally as
$\funceval{\instident{\mix{}}_{\instlang{x86}}}[\instident{pow}_{\instlang{x86}},
\instident{e\!=\!2}] = [\instident{square}_{\instlang{x86}}]$.  The
specialization produces a program that takes a single input (as in
Fig.~\ref{fig:MixExampleOutput} and
$\funceval{\instident{square}_{\instlang{x86}}}[\instident{3}] =
[\instident{9}]$) and squares it. It behaves as a squaring program despite
being comprised of a power program and an input; $\instident{\mix{}}$ has
propagated the input into the original program to produce a specialized
residual program.

\section{The Futamura Projections}
\label{sec:Futamura}

\subsection{First Futamura Projection: Compilation}

\begin{figure}
    \centering
    \begin{subfigure}{\textwidth}
        \centering
        \includegraphics[scale=0.75]{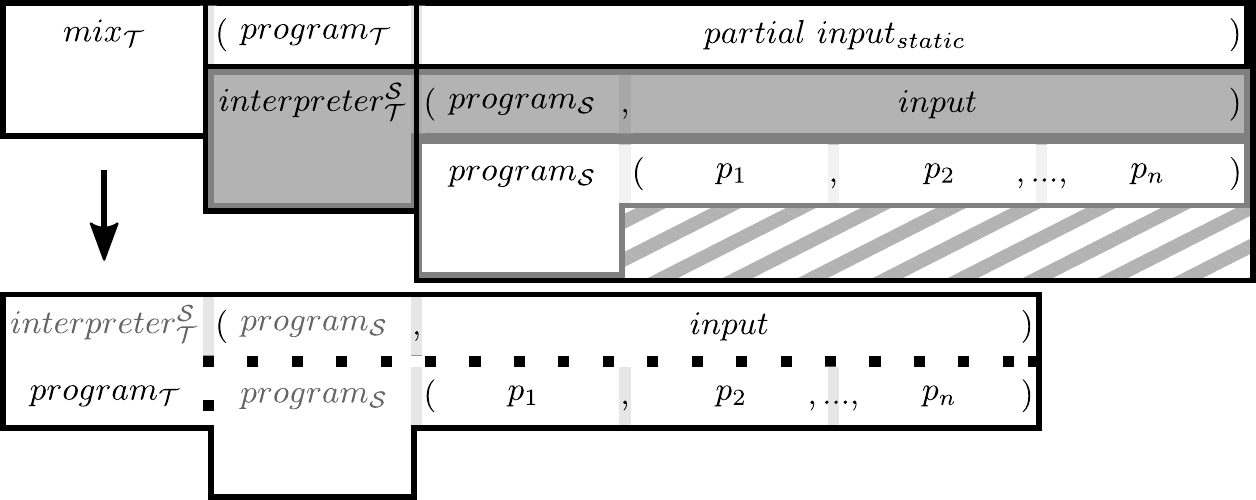}
        \caption{The First Futamura Projection.}
        \label{fig:P1Pattern}
    \end{subfigure}
    \begin{subfigure}{\textwidth}
        \centering
        \includegraphics[scale=0.75]{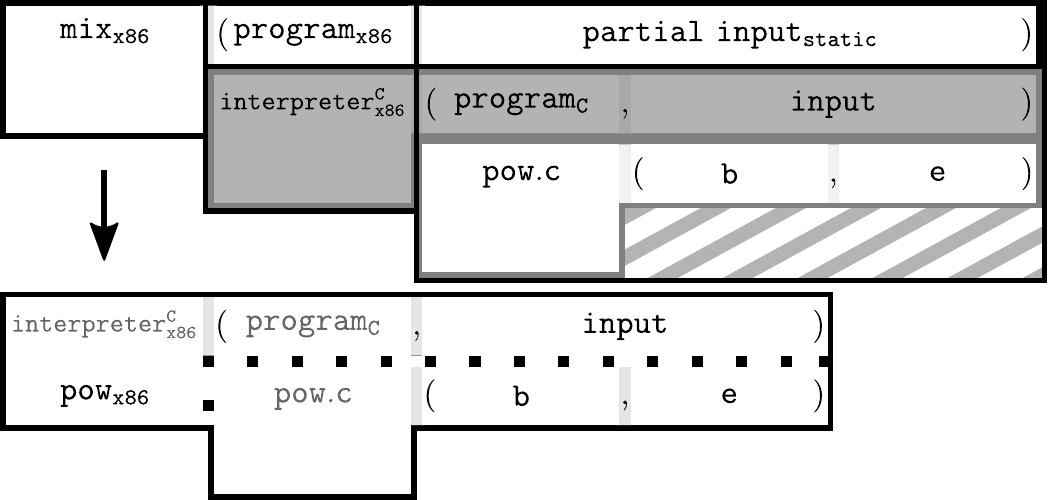}
        \caption{An instance of the First Futamura Projection.}
        \label{fig:P1Example}
    \end{subfigure}
    \begin{subfigure}{\textwidth}
        \centering
        \includegraphics[scale=0.75]{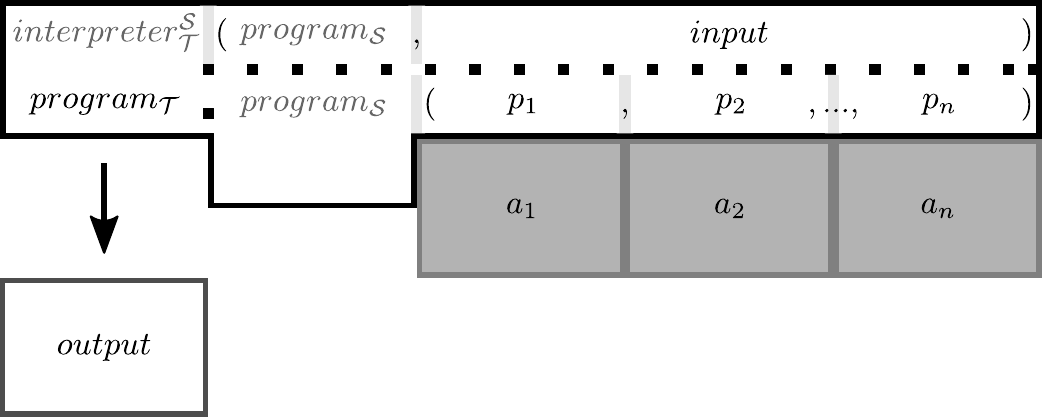}
        \caption{The output of the First Futamura Projection.}
        \label{fig:P1PatternOutput}
    \end{subfigure}
    \begin{subfigure}{\textwidth}
        \centering
        \includegraphics[scale=0.75]{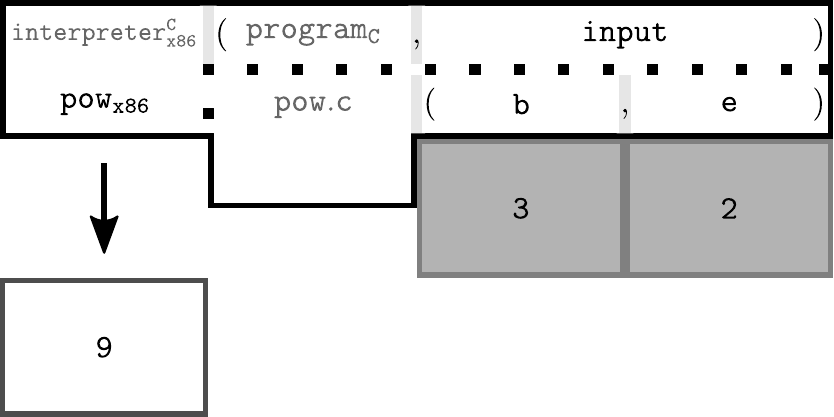}
        \caption{The output of the First Futamura Projection instance.}
        \label{fig:P1ExampleOutput}
    \end{subfigure}
\caption{The First Futamura Projection and example instance.}
\end{figure}

Partial evaluation is beneficial given a program that will be executed repeatedly with some of its input constant, sometimes resulting in a significant speedup. For example, if squaring many values, a specialized squaring program generated from a power program prevents the need for repeated exponent $\instident{e}\!=\!\instident{2}$ arguments. Program interpretation is another case that benefits from partial evaluation; after all, the interpreter is a program and the source program is a subset of its input. Fig.~\ref{fig:P1Pattern} illustrates that when given $\pattident{program}_\pattlang{S}$ and an interpreter for $\pattlang{S}$ implemented in language $\pattlang{T}$, we can partially evaluate the interpreter with the source program as static input
(i.e.,  $\funceval{\pattident{\mix{}}_{\pattlang{T}}}[\pattident{interpreter}^{\pattlang{S}}_{\pattlang{T}}, \pattident{program}_{\pattlang{S}}]
		 = [\pattident{program}_{\pattlang{T}}]$).
This is the \textit{First Futamura Projection}.
As with the previous pattern, the partially evaluated program (i.e., the interpreter) has been specialized to the partial input (i.e., the source program), which is indicated visually by the fusion of the source program to the interpreter.
Notice that $\pattident{program}_\pattlang{S}$ is vertically aligned with the static input slot of the partial evaluator as well as the program input slot of the interpreter.
This is because $\pattident{program}_\pattlang{S}$ serves both roles.
In this case, the dynamic input of the interpreter is the entirety of the input for $\pattident{program}_{\pattlang{S}}$.
When that input is provided in Fig.~\ref{fig:P1PatternOutput}, the specialized program completes the interpretation of $\pattident{program_{\pattlang{S}}}$, producing the output for $\pattident{program_{\pattlang{S}}}$.
\textit{In other words, the specialized program behaves exactly the same as $\pattident{program_{\pattlang{S}}}$, but is implemented in $\pattlang{T}$ rather than $\pattlang{S}$}.
The partial evaluator has effectively \textit{compiled} the program from $\pattlang{S}$ to $\pattlang{T}$.
Thus, the equational form is identical to that of a compiled program:
$\funceval{\pattident{program}_{\pattlang{T}}}[\pattident{a_{1}}, \pattident{a_{2}}, ..., \pattident{a_{n}}] = [\pattident{output}]$.
Fig.~\ref{fig:P1Example} and equation
$\funceval{\instident{\mix{}}_{\instlang{x86}}}[\instident{interpreter}^{\instlang{C}}_{\instlang{x86}}, \instident{pow.c}] = [\instident{pow}_{\instlang{x86}}]$
express the partial evaluation of a C interpreter when given $\instident{pow}$ as partial input. The residual program, detailed in Fig.~\ref{fig:P1ExampleOutput},
behaves the same as $\instident{pow.c}$, but is implemented in \machlang{x86}. The equational expression for the target program is also identical to the compiled program: 
$\funceval{\instident{pow}_{\instlang{x86}}}[\instident{3}, \instident{2}] = [\instident{9}]$.

\begin{quote} \textbf{First Futamura Projection}: A partial evaluator, by
specializing an interpreter to a program, can compile from the interpreted
language to the implementation language of $\instident{\mix{}}$.
\end{quote}

\subsection{Second Futamura Projection: Compiler Generation}

\begin{figure}
    \centering
    \begin{subfigure}{\textwidth}
        \centering
        \includegraphics[scale=0.75]{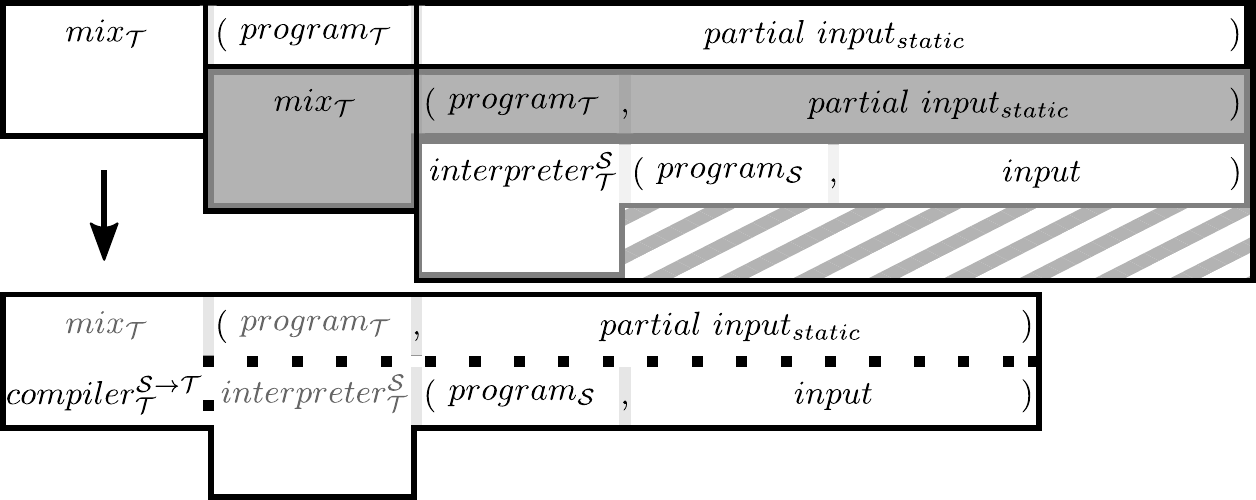}
        \caption{The Second Futamura Projection.}
        \label{fig:P2Pattern}
    \end{subfigure}
    \begin{subfigure}{\textwidth}
        \centering
        \includegraphics[scale=0.75]{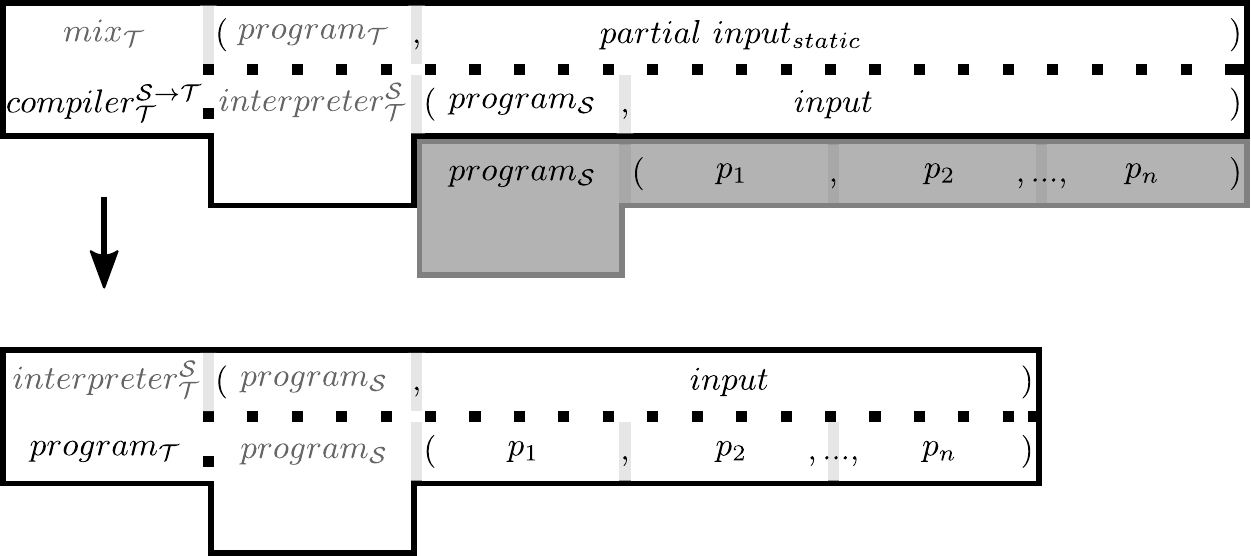}
        \caption{The output of the Second Futamura Projection.}
        \label{fig:P2PatternOutput}
    \end{subfigure}
    \begin{subfigure}{\textwidth}
        \centering
        \includegraphics[scale=0.75]{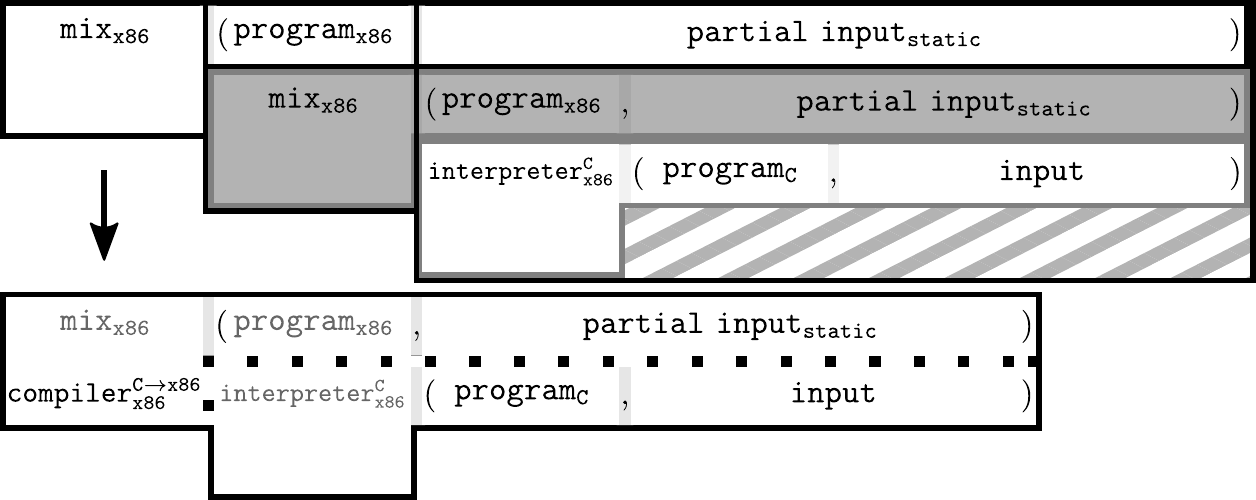}
        \caption{An instance of the Second Futamura Projection.}
        \label{fig:P2Example}
    \end{subfigure}
    \begin{subfigure}{\textwidth}
        \centering
        \includegraphics[scale=0.75]{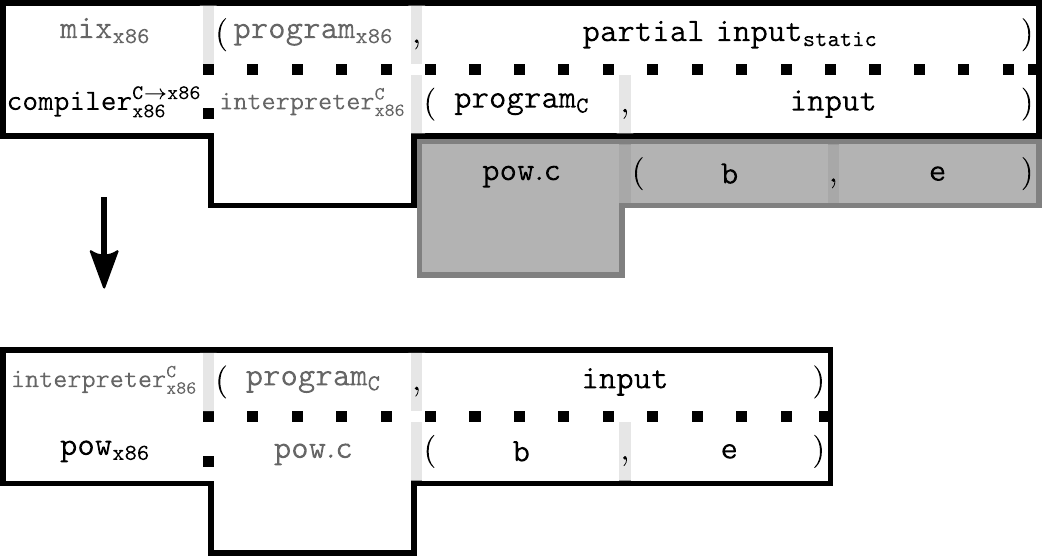}
        \caption{The output of the Second Futamura Projection instance.}
        \label{fig:P2ExampleOutput}
    \end{subfigure}
    \caption{The Second Futamura Projection and example instance.}
\end{figure}

The First Futamura Projection relies on the nature of interpretation requiring two types of input: a program that may be executed multiple times, and input for that program that may vary between executions. As it turns out, the use of $\instident{\mix{}}$ as a compiler exhibits a similar signature: the interpreter is specialized multiple times with different source programs. This allows us to partially evaluate the process of compiling with a partial evaluator. This is the \textit{Second Futamura Projection}, represented equationally as
$\funceval{\pattident{\mix{}}_{\pattlang{T}}}[\pattident{\mix{}}_{\pattlang{T}}, \pattident{interpreter}^{\pattlang{S}}_{\pattlang{T}}]
		 = [\pattident{compiler}^{\pattlang{S}\rightarrow{}\pattlang{T}}_{\pattlang{T}}]$
and depicted in Fig.~\ref{fig:P2Pattern}. In this partial-partial evaluation pattern, an instance of $\instident{\mix{}}$ is being provided as the program input to another instance of $\instident{\mix{}}$, to which an interpreter is provided as static input. Just as in earlier partial evaluation patterns, the program input has been specialized to the given static input; in this case, an instance of $\instident{\mix{}}$ is being specialized to the interpreter. The vertical alignment of programs helps clarify the roles of each program present: the interpreter is the partial input given to the executing instance of $\instident{\mix{}}$ as well as the program input given to the specialized instance of $\instident{\mix{}}$. Additionally, \textit{this specialized residual program as executed in Fig.~\ref{fig:P2PatternOutput} matches the shape and behavior of the First Futamura Projection shown in Fig.~\ref{fig:P1Pattern}}. This is because the same program is being executed with the same input; the only difference is that the output of the second projection is a single program that has been specialized to the interpreter rather than a separate $\instident{\mix{}}$ instance that requires the interpreter to be provided as input. In the Second Futamura Projection, $\instident{\mix{}}$ has generated the $\instident{\mix{}}$-based compiler from the first projection. Because the residual program is a compiler, its equational expression is that of a compiler: 
$\funceval{\pattident{compiler}^{\pattlang{S}\rightarrow{}\pattlang{T}}_{\pattlang{T}}}[\pattident{program}_{\pattlang{S}}]
		 = [\pattident{program}_{\pattlang{T}}]$.

Revisiting the $\instident{pow.c}$ program in Figs.~\ref{fig:P2Example} and \ref{fig:P2ExampleOutput}, $\instident{\mix{}}$ is specialized to a C interpreter to produce a C compiler
(i.e., $\funceval{\instident{\mix{}}_{\instlang{x86}}}[\instident{\mix{}}_{\instlang{x86}}, \instident{interpreter}^{\instlang{C}}_{\instlang{x86}}]
		 = [\instident{compiler}^{\instlang{C}\rightarrow{}\instlang{x86}}_{\instlang{x86}}]$).
When given $\instident{pow}$ in C, this specialized $\instident{\mix{}}$ program then specializes the interpreter to $\instident{pow}$, producing an equivalent power program in \machlang{x86}
($\funceval{\instident{compiler}^{\instlang{C}\rightarrow{}\instlang{x86}}_{\instlang{x86}}}[\instident{pow.c}]
		 = [\instident{pow}_{\instlang{x86}}]$).

\begin{quote} \textbf{Second Futamura Projection}: A partial evaluator, by
specializing another instance of itself to an interpreter, can generate a compiler from the
interpreted language to the implementation language of $\instident{\mix{}}$.
\end{quote}

\subsection{Third Futamura Projection:\\Generation of Compiler Generators}

\begin{figure}
    \centering
    \begin{subfigure}{\textwidth}
        \centering
        \includegraphics[scale=0.75]{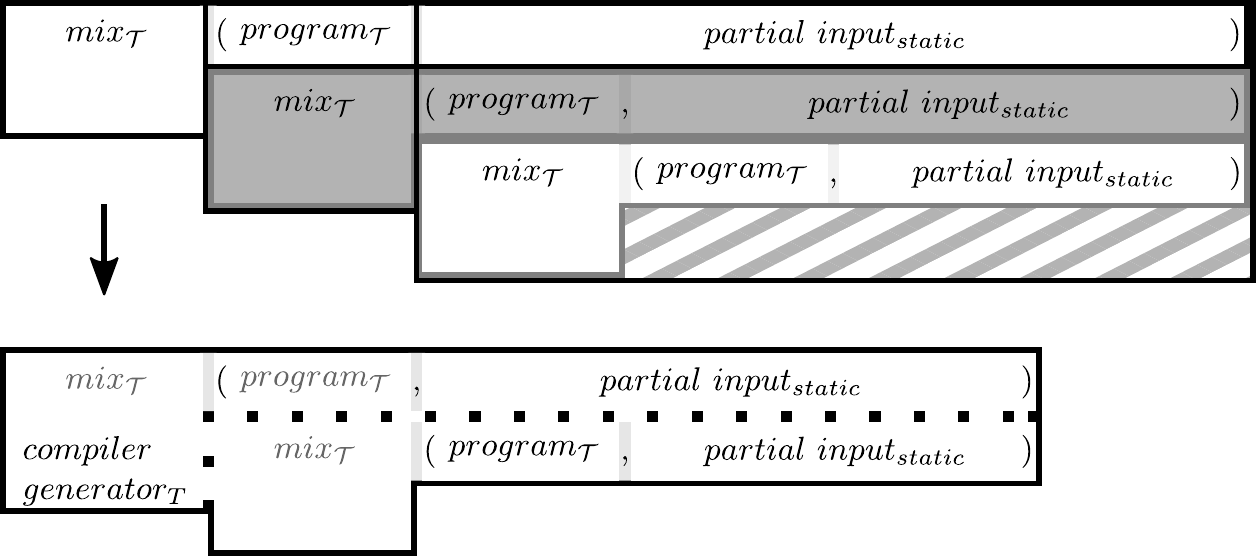}
        \caption{The Third Futamura Projection.}
        \label{fig:P3Pattern}
    \end{subfigure}
    \begin{subfigure}{\textwidth}
        \centering
        \includegraphics[scale=0.75]{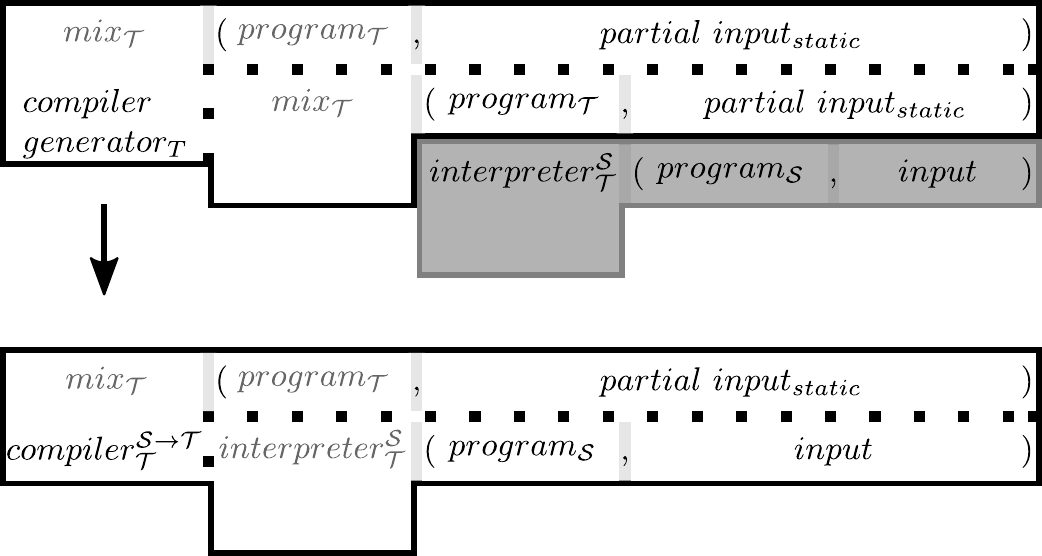}
        \caption{The output of the Third Futamura Projection.}
        \label{fig:P3PatternOutput}
    \end{subfigure}
    \caption{The Third Futamura Projection and output.}
\end{figure}
    
\begin{figure}
    \begin{subfigure}{\textwidth}
        \centering
        \includegraphics[scale=0.75]{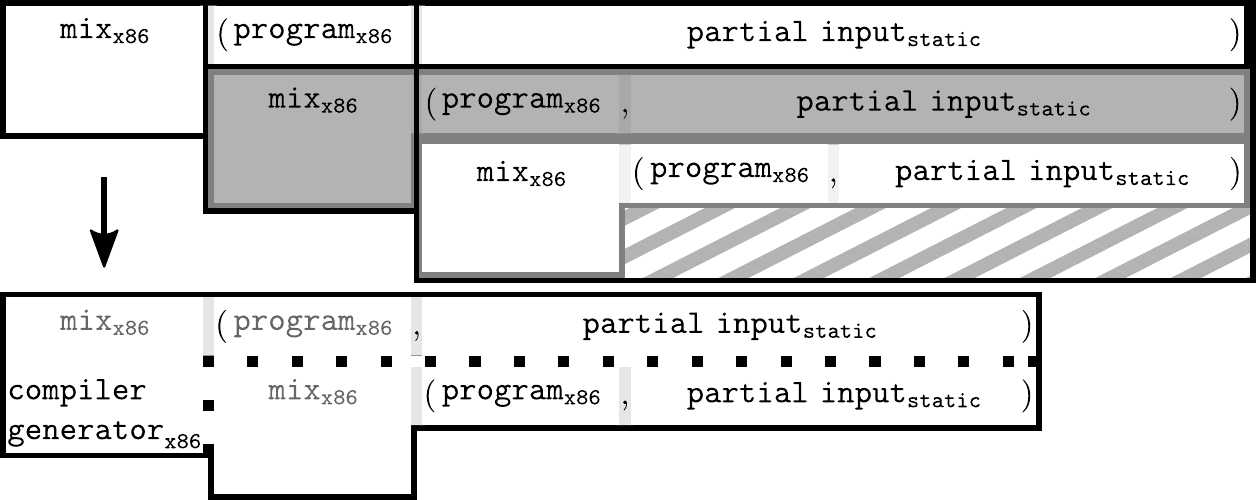}
        \caption{An instance of the Third Futamura Projection.}
        \label{fig:P3Example}
    \end{subfigure}
    \begin{subfigure}{\textwidth}
        \centering
        \includegraphics[scale=0.75]{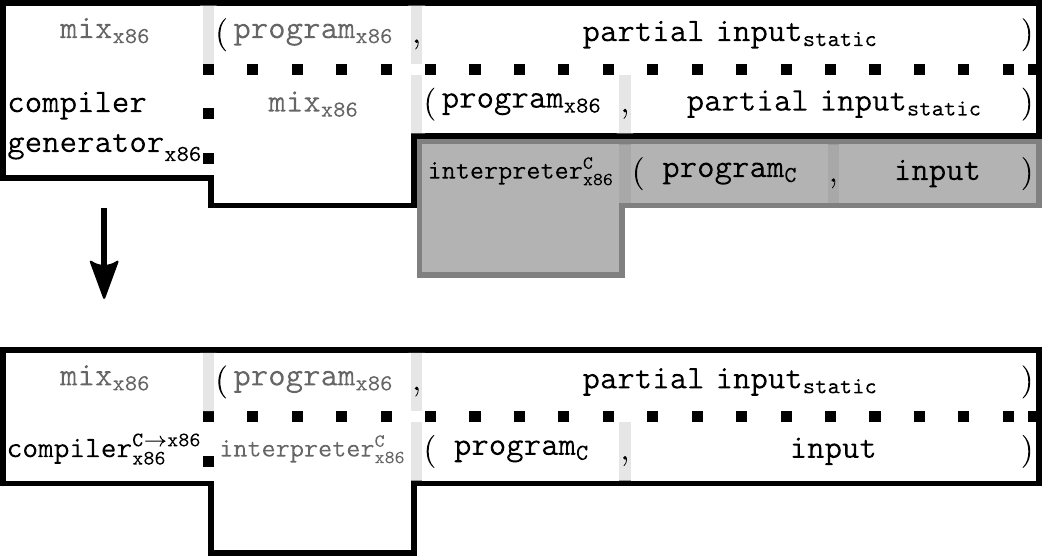}
        \caption{The output of the Third Futamura Projection instance.}
        \label{fig:P3ExampleOutput}
    \end{subfigure}
    \begin{subfigure}{\textwidth}
        \centering
        \includegraphics[scale=0.75]{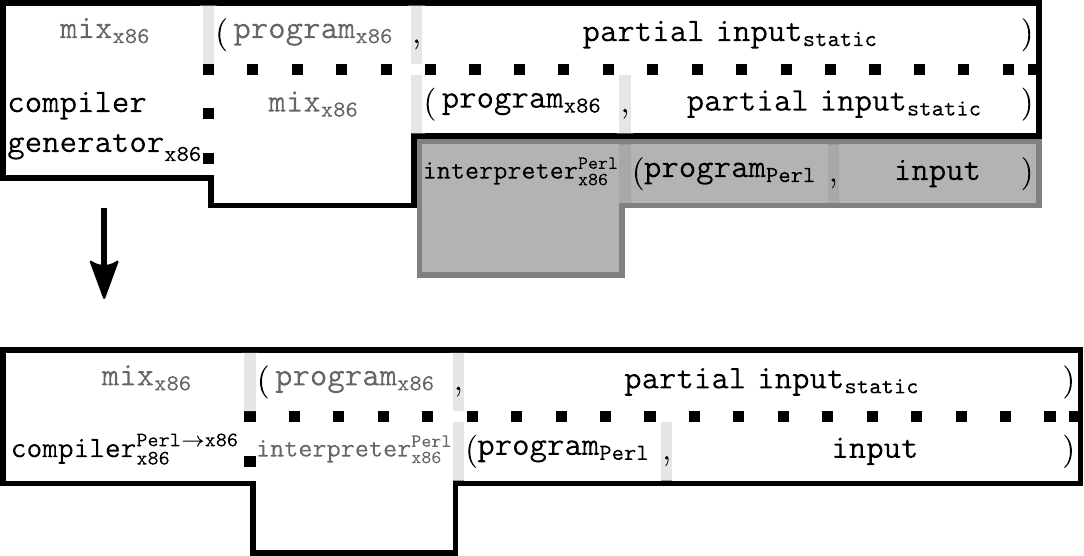}
        \caption{The compiler generator generating a Perl compiler.}
        \label{fig:P3ExampleOutputPerl}
    \end{subfigure}
    \caption{Instance of the Third Futamura Projection and output demonstration.}
\end{figure}
    
Because $\instident{\mix{}}$ can accept itself as input, we can use one instance of $\instident{\mix{}}$ to partially evaluate a second instance of $\instident{\mix{}}$, passing a third instance of $\instident{\mix{}}$ as the static input. This is the \textit{Third Futamura Projection}, shown in Fig.~\ref{fig:P3Pattern} and written equationally as
$\funceval{\pattident{\mix{}}_{\pattlang{T}}}[\pattident{\mix{}}_{\pattlang{T}}, \pattident{\mix{}}_{\pattlang{T}}]
		 = [\pattident{compiler\ generator}_{\pattlang{T}}]$.
The transformation itself is straightforward: partially evaluating a program with some input. The output is still the program in the first input slot specialized to the data in the second input slot; however, this time both the program and the data are instances of $\instident{\mix{}}$. Again, the positioning of the various instances of $\instident{\mix{}}$ within the diagram serves to clarify how the instances interact. The outermost instance executes with the other two instances as input. The middle instance is the ``program'' input of the outer instance and is specialized to the inner instance. Finally, the inner instance is being integrated into the middle instance by the outer instance.
\textit{Notice that Fig.~\ref{fig:P3PatternOutput} shows that the execution of the residual program matches the shape and behavior of the Second Futamura Projection shown in Fig.~\ref{fig:P2Pattern} when provided with an interpreter as input}.
The partial evaluator has generated the $\instident{\mix{}}$-based compiler generator from the second projection. This process is represented equationally
as
$\funceval{\pattident{compiler\ generator}_{\pattlang{T}}}[\pattident{interpreter}^{\pattlang{S}}_{\pattlang{T}}]
		 = [\pattident{compiler}^{\pattlang{S}\rightarrow{}\pattlang{T}}_{\pattlang{T}}]$.

Interestingly, the only variable part of the Third Futamura Projection is the
language associated with $\instident{\mix{}}$.  Previous instance diagrams were
specific to the $\instident{pow.c}$ program; for instance,
Fig.~\ref{fig:P2Example} presents an interpreter for the implementation
language of $\instident{pow.c}$, namely C.  However, the diagram in
Fig.~\ref{fig:P3Example} and the expression
$\funceval{\instident{\mix{}}_{\instlang{x86}}}[\instident{\mix{}}_{\instlang{x86}},
\instident{\mix{}}_{\instlang{x86}}] = [\instident{compiler\
generator}_{\instlang{x86}}]$ make no reference to $\instident{pow.c}$ or C.
This is because the Third Futamura Projection has abstracted the interpretation
process to an extent that even the interpreter is considered dynamic input.
Fig.~\ref{fig:P3ExampleOutput} shows the $\instident{\mix{}}$-generated
compiler generator accepting a C interpreter and generating a C to
\machlang{x86} compiler ($\funceval{\instident{compiler\
generator}_{\instlang{x86}}}[\instident{interpreter}^{\instlang{C}}_{\instlang{x86}}]
=
[\instident{compiler}^{\instlang{C}\rightarrow{}\instlang{x86}}_{\instlang{x86}}]$),
but it will accept \textit{any} interpreter implemented in \machlang{x86}
regardless of the language interpreted. For example,
Fig.~\ref{fig:P3ExampleOutputPerl} shows a compiler for the language Perl being
generated by the same compiler generator (i.e., $\funceval{\instident{compiler\
generator}_{\instlang{x86}}}[\instident{interpreter}^{\instlang{Perl}}_{\instlang{x86}}]
=
[\instident{compiler}^{\instlang{Perl}\rightarrow{}\instlang{x86}}_{\instlang{x86}}]$).
A residual compiler generated through the result of the Third Futamura
Projection is called a \textit{generating extension}---a term coined by
Ershov~\cite{ershovPE}---of the input interpreter~\cite{cogenSixLines}. In
general, the result of The Third Projection creates a generating extension for
any program provided to it.
         
\begin{quote}
\textbf{Third Futamura Projection}: A partial evaluator, by specializing an additional
instance of itself to a third instance, can generate a compiler generator that produces
compilers from any language to the implementation language of
$\instident{\mix{}}$.
\end{quote}
    
\subsection{Summary: Futamura Projections}

The Third Futamura Projection follows the pattern of the previous two
projections: the use of $\instident{\mix{}}$ to partially evaluate a prior process (i.e., interpretation, compilation).
The first projection compiles by partially evaluating the interpretation
process without the input of the source program. The Second Futamura Projection generates a
compiler by partially evaluating the compilation process of the first
projection without any particular source program. The Third Futamura Projection generates
a compiler generator by partially evaluating the compiler generation process of the second projection
without an interpreter. Each projection delays completion of the previous process
by abstracting away the more variable of two inputs. Just as the first
projection interprets a program with various, dynamic inputs and the
second projection compiles various programs, the third projection generates
compilers for various languages/interpreters.  Table~\ref{tab:Summary}
juxtaposes the related equations and diagrams from both $\S$~\ref{sec:Processing}
and $\S$~\ref{sec:Futamura} in each row to make their relationships more
explicit.  Each row of Table~\ref{tab:ProjectionSummary} succinctly summarizes
each projection by associating each side of its equational representation with the corresponding
diagram from $\S$~\ref{sec:Futamura}.

\begin{sidewaystable}
\caption{Juxtaposition of related equations and
diagrams from $\S$~\ref{sec:Processing} and $\S$~\ref{sec:Futamura}.}
\resizebox{\textwidth}{!}{
\begin{tabular}{|l|l|l|l|l|l|}
\hline
\multicolumn{1}{|c|}{\multirow{2}{*}{\textbf{Fig.}}} & \multicolumn{1}{c|}{\multirow{2}{*}{\textbf{Equational Notation}}}
    & \multicolumn{1}{c|}{\multirow{2}{*}{\textbf{Input Description}}} & \multicolumn{1}{c|}{\multirow{2}{*}{\textbf{Output Description}}}
    & \multicolumn{1}{c|}{\textbf{Similar}} & \multicolumn{1}{c|}{\textbf{Output}} \\
    & & &
    & \multicolumn{1}{c|}{\textbf{ Fig.}} & \multicolumn{1}{c|}{\textbf{ Fig.}} \\
\hline
\ref{fig:BasicPattern} & $\funceval{\pattident{program}}[\pattident{a_{1}}, \pattident{a_{2}}, ..., \pattident{a_{n}}] = [\pattident{output}]$
    & A list of arguments to the program. & The result of the program's execution. & \multicolumn{1}{|c|}{N/A} & \multicolumn{1}{|c|}{N/A} \\
\hline
\ref{fig:CompilerPattern}
    & $\funceval{\pattident{compiler}^{\pattlang{S}\rightarrow{}\pattlang{T}}_{\pattlang{T}}}[\pattident{program}_{\pattlang{S}}] = [\pattident{program}_{\pattlang{T}}]$
    & A program in language $\pattlang{S}$. & The input program in language $\pattlang{T}$. & \multicolumn{1}{|c|}{N/A} & \multicolumn{1}{|c|}{\ref{fig:CompilerPatternOutput}} \\
\hline
\ref{fig:InterpreterPattern}
    & $\funceval{\pattident{interpreter}^{\pattlang{S}}_{\pattlang{T}}}[\pattident{program}_{\pattlang{S}}, \pattident{input}] = [\pattident{output}]$
    & A program in language $\pattlang{S}$, along with its input. & The result of the program's execution. & \multicolumn{1}{|c|}{N/A} & \multicolumn{1}{|c|}{N/A} \\
\hline

\ref{fig:MixPattern}
    & $\funceval{\pattident{\mix{}}_{\pattlang{T}}}[\pattident{program}_{\pattlang{T}}, \pattident{partial\ input_{static}}] = [\pattident{program^{\prime}}_{\pattlang{T}}]$
    & A program in language $\pattlang{T}$, along with any subset of its input. & The input program specialized to the partial input. & \multicolumn{1}{|c|}{N/A} & \multicolumn{1}{|c|}{\ref{fig:MixPatternOutput}} \\
\hline
\ref{fig:MixPatternOutput}
    & $\funceval{\pattident{program^{\prime}}_{\pattlang{T}}}[\pattident{a_{1}}, \pattident{a_{3}}, ..., \pattident{a_{n}}] = [\pattident{output}]$
    & A list of arguments to the original program, excluding the static input. & The result of the original program's execution. & \multicolumn{1}{|c|}{\ref{fig:BasicPattern}} & \multicolumn{1}{|c|}{N/A} \\
\hline

\ref{fig:P1Pattern}
    & $\funceval{\pattident{\mix{}}_{\pattlang{T}}}[\pattident{interpreter}^{\pattlang{S}}_{\pattlang{T}}, \pattident{program}_{\pattlang{S}}]
            = [\pattident{program}_{\pattlang{T}}]$
    & An interpreter for language $\pattlang{S}$ and a program implemented in $\pattlang{S}$. & The program implemented in $\pattlang{T}$. & \multicolumn{1}{|c|}{N/A} & \multicolumn{1}{|c|}{\ref{fig:P1PatternOutput}} \\
\hline
\ref{fig:P1PatternOutput}
    & $\funceval{\pattident{program}_{\pattlang{T}}}[\pattident{a_{1}}, \pattident{a_{2}}, ..., \pattident{a_{n}}]
            = [\pattident{output}]$
    & A list of arguments to the original program. & The result of the original program's execution. & \multicolumn{1}{|c|}{\ref{fig:InterpreterPattern}} & \multicolumn{1}{|c|}{N/A} \\
\hline

\ref{fig:P2Pattern}
    & $\funceval{\pattident{\mix{}}_{\pattlang{T}}}[\pattident{\mix{}}_{\pattlang{T}}, \pattident{interpreter}^{\pattlang{S}}_{\pattlang{T}}]
            = [\pattident{compiler}^{\pattlang{S}\rightarrow{}\pattlang{T}}_{\pattlang{T}}]$
    & $\pattident{\mix{}}$ and an interpreter for language $\pattlang{S}$, both implemented in $\pattlang{T}$. & A compiler from $\pattlang{S}$ to $\pattlang{T} $
    & \multicolumn{1}{|c|}{N/A} & \multicolumn{1}{|c|}{\ref{fig:P2PatternOutput}} \\
\hline
\ref{fig:P2PatternOutput}
    & $\funceval{\pattident{compiler}^{\pattlang{S}\rightarrow{}\pattlang{T}}_{\pattlang{T}}}[\pattident{program}_{\pattlang{S}}]
            = [\pattident{program}_{\pattlang{T}}]$
    & A program in language $\pattlang{S}$. & The input program in language $\pattlang{T}$.
    & \multicolumn{1}{|c|}{\ref{fig:P1Pattern}} & \multicolumn{1}{|c|}{\ref{fig:P1PatternOutput}} \\
\hline

\ref{fig:P3Pattern}
    & $\funceval{\pattident{\mix{}}_{\pattlang{T}}}[\pattident{\mix{}}_{\pattlang{T}}, \pattident{\mix{}}_{\pattlang{T}}]
            = [\pattident{compiler\ generator}_{\pattlang{T}}]$
    & Two instances of $\pattident{\mix{}}$ implemented in $\pattlang{T}$. & A compiler generator implemented in $\pattlang{T}$ & \multicolumn{1}{|c|}{N/A} & \multicolumn{1}{|c|}{\ref{fig:P3PatternOutput}} \\
\hline
\ref{fig:P3PatternOutput}
    & $\funceval{\pattident{compiler\ generator}_{\pattlang{T}}}[\pattident{interpreter}^{\pattlang{S}}_{\pattlang{T}}]
            = [\pattident{compiler}^{\pattlang{S}\rightarrow{}\pattlang{T}}_{\pattlang{T}}]$
    & An interpreter for language $\pattlang{S}$. & A compiler from $\pattlang{S}$ to $\pattlang{T}.$ & \multicolumn{1}{|c|}{\ref{fig:P2Pattern}} & \multicolumn{1}{|c|}{\ref{fig:P2PatternOutput}} \\

\hline
\end{tabular}}
\label{tab:Summary}

\caption{Summary of Futamura Projections.}
\resizebox{\textwidth}{!}{
\begin{tabular}{|l|l|l|l|l|l|}
\hline
\multicolumn{1}{|c|}{\textbf{Projection}} & \multicolumn{1}{c|}{\textbf{Description}}
    & \multicolumn{1}{c|}{\textbf{Equational Notation}}
    & \multicolumn{1}{c|}{\textbf{Fig.}} & \multicolumn{1}{c|}{\textbf{ Output Fig.}} \\
\hline

\multicolumn{1}{|c|}{1} & $\instident{mix}$ can compile.
    & $\funceval{\pattident{\mix{}}_{\pattlang{T}}}[\pattident{interpreter}^{\pattlang{S}}_{\pattlang{T}}, \pattident{program}_{\pattlang{S}}]
                    = [\pattident{program}_{\pattlang{T}}]$
    & \ref{fig:P1Pattern} & \multicolumn{1}{|c|}{\ref{fig:P1PatternOutput}} \\
\hline

\multicolumn{1}{|c|}{2} & $\instident{mix}$ can generate a compiler.
    & $\funceval{\pattident{\mix{}}_{\pattlang{T}}}[\pattident{\mix{}}_{\pattlang{T}}, \pattident{interpreter}^{\pattlang{S}}_{\pattlang{T}}]
            = [\pattident{compiler}^{\pattlang{S}\rightarrow{}\pattlang{T}}_{\pattlang{T}}]$
    & \ref{fig:P2Pattern} & \multicolumn{1}{|c|}{\ref{fig:P2PatternOutput}} \\
\hline

\multicolumn{1}{|c|}{3} & $\instident{mix}$ can generate a compiler generator.
    & $\funceval{\pattident{\mix{}}_{\pattlang{T}}}[\pattident{\mix{}}_{\pattlang{T}}, \pattident{\mix{}}_{\pattlang{T}}]
            = [\pattident{compiler\ generator}_{\pattlang{T}}]$
    & \ref{fig:P3Pattern} & \multicolumn{1}{|c|}{\ref{fig:P3PatternOutput}} \\

\hline
\end{tabular}}
\label{tab:ProjectionSummary}
\end{sidewaystable}

\section{Discussion}

\subsection{Examples of Partial Evaluators}

Partial evaluators have been developed for several practical programming
languages including C~\cite{mixForC}, Scheme~\cite{similix}, and
Prolog~\cite{logimix}.  These partial evaluators are self-applicable and, thus,
can perform all three Futamura Projections~\cite{fourthProj}. However concerns
such as performance make the projections impractical with these programs.

\subsection{Beyond the Third Projection}

Researchers have studied what lies beyond the third projection and what
significance the presence of a fourth projection might have~\cite{fourthProj}.
Two important conclusions have been made in this regard:

\begin{itemize}

\item Any self-generating compiler generator, i.e., a compiler generator such that,

\begin{displaymath}
\funceval{compiler\ generator}[mix] = compiler\ generator\mathrm{,}
\end{displaymath}

\noindent can be obtained by repeated self-application of a partial evaluator
as in the Third Futamura Projection, and vice versa.

\item The compiler generator can be applied to another partial evaluator with different properties to produce a new compiler generator with related properties. For example, applying a compiler generator that accepts C programs to a partial evaluator accepting Python (but written in C) will produce a compiler generator that accepts Python programs.

\end{itemize}

\noindent These two observations are not orthogonal, but rather two sides of
the same coin because they both depend on the application of a 
compiler generator to a partial evaluator.
We refer the reader interested in further, formal
exploration of these ideas and insights to~\cite{fourthProj}.

\subsection{Applications}

\subsubsection{Truffle and Graal}

\begin{figure}
	\centering
	\includegraphics[scale=0.75]{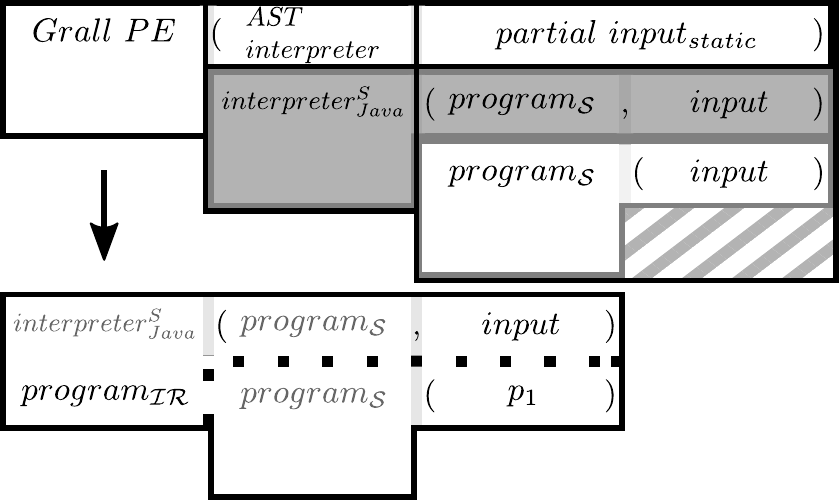}
	\caption{The First Futamura Projection inside the Graal JIT.}
	\label{fig:GraalPattern}
\end{figure}

The \textit{Truffle} project, developed by Oracle Labs, seeks to facilitate the
implementation of fast, dynamic languages. It provides a framework of Java
classes and annotations that allows language developers to build abstract
syntax tree interpreters and indicate ways in which program behavior may be
optimized~\cite{Truffle}. When such an interpreter is applied to a program
through the \textit{Graal} compilation infrastructure, the Graal just-in-time
compiler (JIT) compiles program code to a custom intermediate representation by
applying a partial evaluator at run-time~\cite{Graal}.  We illustrate this JIT
compilation using our diagram notation in Fig.~\ref{fig:GraalPattern}. This is
a special form of the First Futamura Projection where only certain fragments of
the source program are compiled based on suggestions communicated through the
Truffle Domain Specific Language.  Through partial evaluation, Truffle and
Graal provide a language implementation option that leverages the established
benefits of the host virtual machine such as tool support, language
interoperability, and memory management.

Because the Graal partial evaluator both specializes and is written in Java, it
could in theory be used to reach the second and third projections. However, it
was designed specifically to specialize interpreters written with Truffle and
makes optimization assumptions about the execution of the program. As a result,
the partial evaluator is not practically self-applicable and cannot efficiently
be used for the second projection.

\subsubsection{PyPy}

\begin{figure}
\begin{subfigure}{0.49\textwidth}
	\centering
	\includegraphics[width=\textwidth]{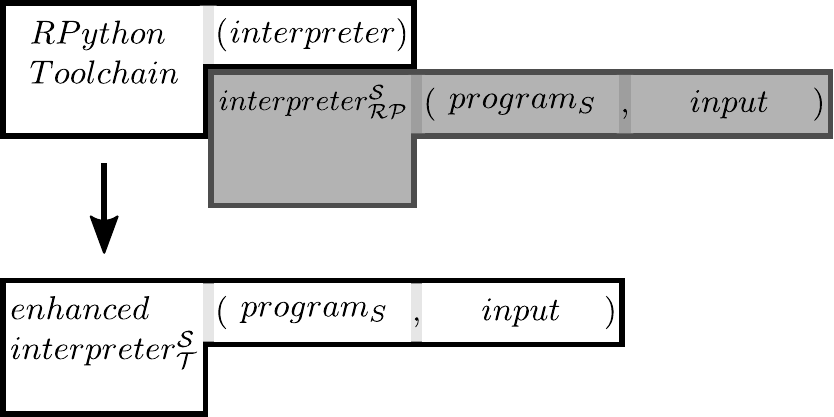}
	\caption{The RPython transformation of an interpreter.}
	\label{fig:RPythonPattern}
\end{subfigure}
\begin{subfigure}{0.49\textwidth}
	\centering
	\includegraphics[width=\textwidth]{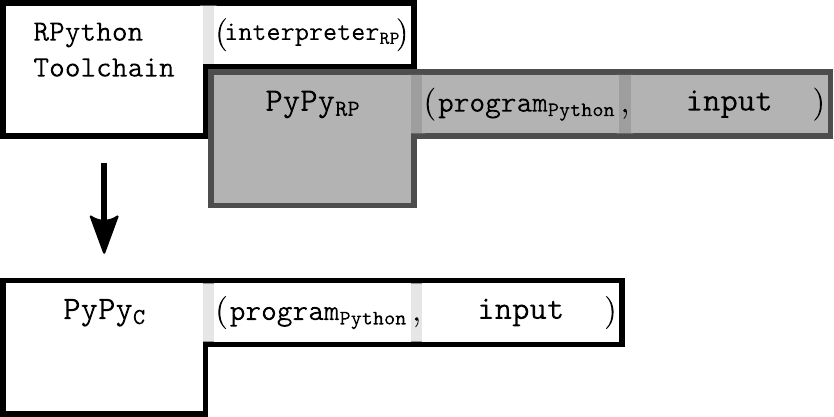}
	\caption{The PyPy self-interpreter, compiled to C.}
	\label{fig:RPythonExample}
\end{subfigure}
\caption{The PyPy project, represented in our diagram notation.}
\end{figure}

We would be remiss not to mention the \textit{PyPy} project
(\url{https://pypy.org/}), which approaches the problem of optimizing
interpreted programs from a different perspective. While named for its Python
self-interpreter, the pertinent part of the project is its RPython translation
toolchain. RPython is a framework designed for compiling high-level, dynamic
language interpreters from a subset of Python to a selection of low-level
languages such as C or bytecode in a way that adds features common to virtual
machines such as memory management and JIT compilation.  To demonstrate this,
the PyPy self-interpreter, once compiled to C, outperforms the CPython
interpreter in many performance tests~\cite{PyPySpeed}.  The translation
process is depicted using our diagram scheme in Fig.~\ref{fig:RPythonPattern},
with the application to the PyPy interpreter in Fig.~\ref{fig:RPythonExample}.

Although both the Truffle/Graal and PyPy projects make use of JIT compilation
for optimization purposes, PyPy's JIT is different from that used by Graal.
While both compile the source program indirectly through the interpreter, only
Graal performs the First Futamura Projection by compiling through partial
evaluation. PyPy, on the other hand, uses a \textit{Tracing JIT compiler} that
traces frequently executed code and caches the compiled version to bypass
interpretation~\cite{PyPy}.  For a more in-depth comparison of the methods used
by Truffle/Graal and PyPy, including performance measurements, we refer the
reader to~\cite{PTGCompare}.

\subsection{Conclusion}

The Futamura Projections can be used to compile, generate compilers, and
generate compiler generators.  We are optimistic that this article has
demystified their esoteric nature.  Increased attention for and broader
awareness of this topic may lead to varied perspectives on
language implementation and optimization tools like
Truffle/Graal and PyPy.  Additionally, further analysis may lead
to new strides into the development of a practical partial evaluator that can
effectively produce the Futamura Projections.

\bibliographystyle{alphaurl}
\bibliography{futamura}
\end{document}